\documentclass{scienceadvhack}

\usepackage{scicite}
\usepackage{times}
\usepackage{graphicx}
\usepackage{amsmath}
\usepackage{amssymb}

\usepackage{newfloat}
\DeclareFloatingEnvironment[name=Extended Figure]{extFigure}

\newcommand{\tess}{{\it TESS}}

\newcounter{lastnote}

\title{An accreting white dwarf displaying fast transitional mode switching }

\author
{Simone Scaringi$^{1\ast}$,
Domitilla de Martino$^{2}$,
David A.H. Buckley$^{3,4,5}$,
Paul J. Groot$^{6,3,4}$, 
Christian Knigge$^{7}$,
Matteo Fratta$^{1}$,
Krystian Ilkiewicz$^{1}$,
Colin Littlefield$^{8,9}$,
Alessandro Papitto$^{10}$. 
\\
\\
\normalsize{$^{1}$Centre for Extragalactic Astronomy, Department of Physics, Durham University, DH1 3LE, UK}\\
\normalsize{$^{2}$INAF-Osservatorio Astronomico di Capodimonte, Salita Moiariello 16, I-80131 Naples, Italy}\\
\normalsize{$^{3}$South African Astronomical Observatory,  PO Box 9, Observatory, 7935, Cape Town, South Africa}\\
\normalsize{$^{4}$Department of Astronomy, University of Cape Town, Private Bag X3, Rondebosch, 7701, South Africa}\\
\normalsize{$^{5}$Department of Physics, University of the Free State, PO Box 339, Bloemfontein, 9300, South Africa}\\
\normalsize{$^{6}$Department of Astrophysics/IMAPP, Radboud University, P.O. 9010, 6500 GL, Nĳmegen, The Netherlands}\\
\normalsize{$^{7}$School of Physics and Astronomy, University of Southampton, Highfield, Southampton SO17 1BJ, UK}\\
\normalsize{$^{8}$Department of Physics, University of Notre Dame, Notre Dame, IN 46556, USA}\\
\normalsize{$^{9}$Department of Astronomy, University of Washington, Seattle, WA 98195, USA}\\
\normalsize{$^{10}$INAF-Osservatorio Astronomico di Roma, Via Frascati 33, I-00078, Monteporzio Catone (Rome), Italy}\\
\\
\normalsize{$^\ast$Corresponding author. E-mail: simone.scaringi@durham.ac.uk}
\\
\normalsize{Submitted on 17 June 2021. Accepted for publication in \textit{Nature Astronomy} on 18 August 2021.}
}

\date{}

\begin{document}

\maketitle 

\begin{abstract}
Accreting white dwarfs are often found in close binary systems with orbital periods ranging from tens of minutes to several hours. In most cases, the accretion process is relatively steady, with significant modulations only occurring on time-scales of $\approx$days or longer$^{1,2}$. Here, we report the discovery of abrupt drops in the optical luminosity of the accreting white dwarf binary system TW Pictoris by factors up to 3.5 on time-scales as short as 30 minutes. The optical light curve of this binary system obtained by the \textit{Transiting Exoplanet Survey Satellite} (\tess) clearly displays fast switches between two distinct intensity modes that likely track the changing mass accretion rate onto the white dwarf. In the low mode, the system also displays magnetically-gated accretion bursts$^{3,4,5}$, which implies that a weak magnetic field of the white dwarf truncates the inner disk at the co-rotation radius in this mode. The properties of the mode switching observed in TW Pictoris appear analogous to those observed in transitional millisecond pulsars$^{6,7,8,9,10}$, where similar transitions occur, although on timescales of $\approx$tens of seconds. Our discovery establishes a previously unrecognised phenomenon in accreting white dwarfs and suggests a tight link to the physics governing magnetic accretion onto neutron stars.
\end{abstract}

The binary system TW Pictoris (hereafter TW Pic) is an accreting white dwarf first identified as the optical counterpart of the High Energy Astronomy Observatory (HEAO-1) X-ray source H0534$-$58111. It is usually observed in an optically bright ($m_V\approx 14$), high-luminosity state, but is known to occasionally drop in flux by more than a factor of $\approx$3 at both X-ray and optical wavelengths$^{11,12}$. The exact classification of TW Pic has been controversial since its discovery. Optical spectroscopy clearly reveals the presence of the high-ionisation HeII 4686\AA\ line, as well as the usual Balmer series and HeI lines observed in essentially all accreting white dwarfs. The presence of HeII is generally associated with systems with an intense high energy ionizing source. More specifically, systems harbouring magnetic white dwarfs can channel material onto the magnetic polar regions where a stand-off shock is formed, below which X-ray radiation is emitted13. In most of these systems the magnetic field of the white dwarf is strong enough ($B>10^6$G at the white dwarf surface) to disrupt the inner accretion disk. Due to the asynchronous fast rotation of the white dwarf relative to the binary orbital period, these systems generally display periodic signals at the white dwarf spin, binary orbit, and/or orbital sidebands$^{14}$, and are generally referred to as intermediate polars. Systems with stronger magnetic white dwarfs ($B>10^7$ G) , referred to as polars, can entirely prevent the formation of an accretion disk, and the white dwarf rotates synchronously at the binary orbital period. X-ray and optical modulations are generally observed for all these magnetic systems. Although there have been claims of detected spin and orbital signals in optical observations$^{15}$ of TW Pic, and the former also in X-rays$^{16}$, they were not found to be persistent, and sometimes variable$^{12}$, thus casting doubts on an intermediate polar nature of TW Pic.

TW Pic has been observed with the \tess\ satellite during Cycle 1 (from 25 July 2018 to 17 July 2019) at 120 sec cadence, and in Cycle 3 (from 5 July 2020 to 28 April 2021) at 20 sec cadence. \tess\ operates over the wide passband of 600-1000nm. A transient signal at $P\approx 6.956$ hours is detected during Cycle 1 and possibly also during Cycle 3 (see Methods and Extended Data Fig.~1), which may be related to the orbit-induced signals of the system. More surprising, the \tess\ observations during Cycle 3 (Fig.~1) display abrupt and sudden switches in the system luminosity, dropping by more than a factor of 3.5 on timescales shorter than $\approx$30 minutes. This moding has happened at least 4 times during the \tess\ Cycle 3 observations with variable durations of the low mode from $\approx$1 day to $\approx$2 months (see Extended Data Fig.~2). Upon closer inspection, g-band data from the All-Sky Automated Survey for Supernovae$^{17,18}$ (ASAS-SN) indicates that this moding behaviour has also happened between September 2019 and March 2020 when \tess\ was not observing the source (see Extended Data Fig.~3). Although the ASAS-SN lightcurve indicates that TW Pic can undergo slow transitions from a high to a low state (and vice versa) on timescales of weeks to months as is observed in several other accreting white dwarfs$^{1,19,20}$, the moding behaviour occurring on $\approx$30 minute timescales appears to be restricted to the more recent observations. The \tess\ Cycle 3 data also reveal the mode switching from the low to the high mode, occurring on a somewhat longer timescale of $\approx$12 hours (see Extended Data Fig.~2). These transitions appear to be remarkably reproducible. From close visual inspection of both Cycle 1 and Cycle 3 \tess\ data it is clear that TW Pic attempted to abruptly enter the low mode on several occasions on the same $\approx$30 minute timescales, but failed to maintain the low luminosity levels and instead rapidly recovered to the high mode again (see Extended Data Fig.~4).

\begin{figure*}[h]
\begin{center}
\includegraphics[width = \textwidth]{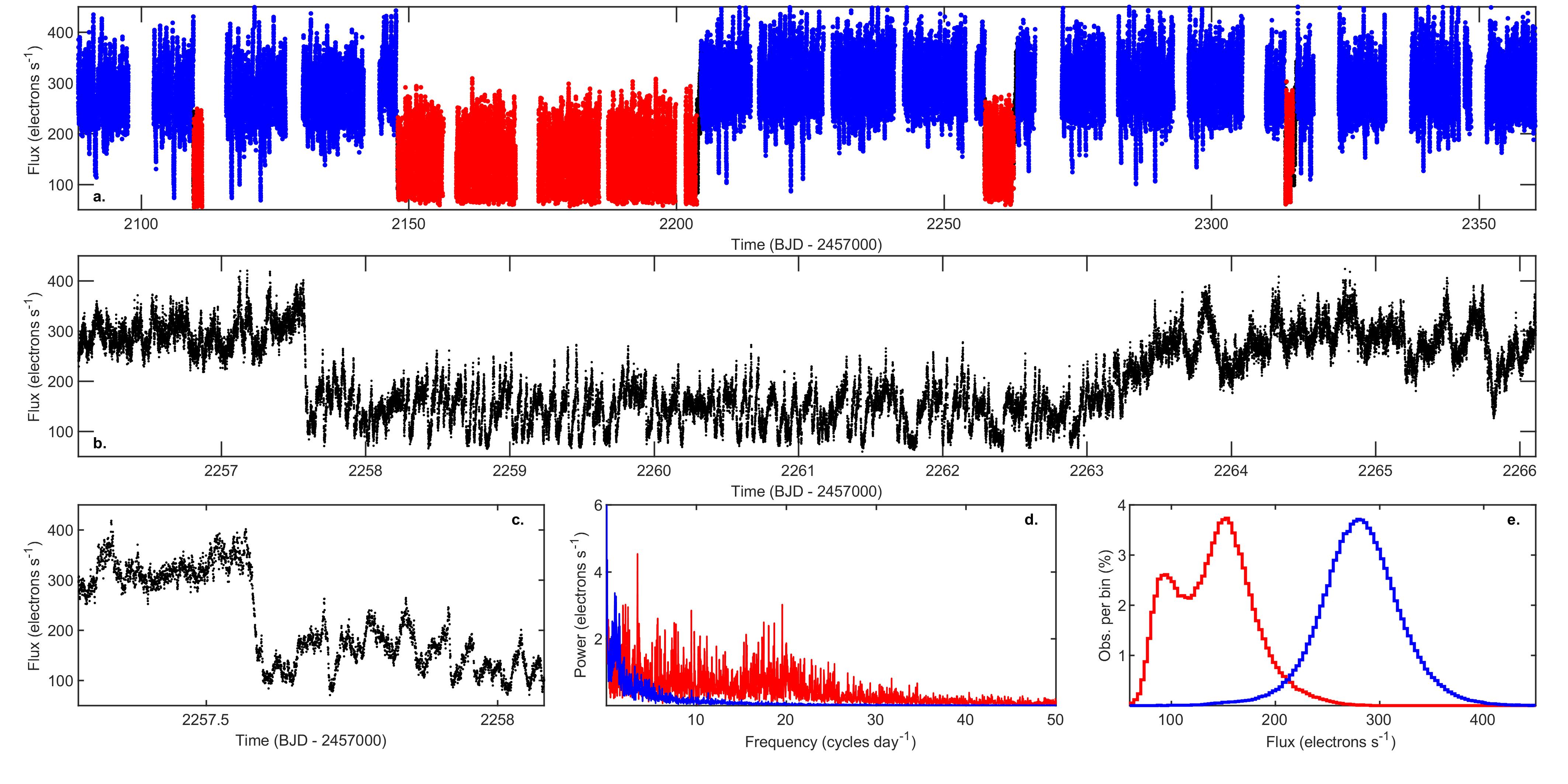}
\caption{
\noindent \textbf{Optical brightness variations in TW Pic observed with \tess.}
\textbf{a.} \tess\ lightcurve (20 sec cadence) between 26 August 2020 and 28 April 2021 (Sectors 27 - 38). Marked with blue data points are the selected high modes, while in red we mark the low modes. The transitions are marked in black. \textbf{b.} Zoom of the \tess\ lightcurve during the continuous observation of both mode switches. The inset reveals the magnetically gated bursts of duration $\approx$30 minutes every $\approx$1.2-2.4 hours overlaid onto a constant (flat) luminosity level. \textbf{c.} Further zoom of the \tess\ lightcurve displaying a rapid $\approx$30 minute mode switch. \textbf{d.} Lomb-scargle periodogram during the high mode (blue) and low mode (red). The excess variability power between 10-20 cycles/day can be attributed to the magnetically gated bursts. \textbf{e.} Histogram of \tess\ recorded fluxes in the low (red) and high (blue) mode clearly showing the different modes. The bimodality shown in the low mode is attributed to the magnetically gated bursts.
}
\end{center}
\end{figure*}

During the low mode, TW Pic displays clear bursts with flux variations by a factor of $\approx$2 recurring on timescales of $\approx$1.2-2.4 hours. This variability can be attributed to magnetically gated bursts, only observed in one other accreting white dwarf to date, namely MV Lyrae$^{3}$, which however did not display the mode switches as observed in TW Pic. These quasi-periodic bursts in optical light are thought to occur at very low mass-transfer rates$^{3,21}$ and are caused by residual disk material overcoming the spinning magnetospheric barrier in discrete episodic accretion events. Their presence allows us to constrain the accretion rate onto the white dwarf during the low mode. To do this we transformed the \tess\ lightcurve of TW Pic into g-band magnitudes using the archival simultaneous observations obtained by ASAS-SN (see Methods). We estimate the average high mode level to be $m_g\approx 14.2$ and the lowest brightness level to be $m_g\approx 16.9$ attained by TW Pic during Cycle 3 observations. The latter is consistent with previous ASAS-SN observations of TW Pic in the low state. This brightness level is, however, not the lowest observed in this system (see Methods), indicating that a residual contribution from an accretion disk is still present in the low mode. Using the time-averaged magnitude during the whole low-mode period of $m_g\approx 15.5$, which includes the observed quasi-periodic bursts, this translates to a time-averaged mass accretion rate onto the white dwarf of $\gtrsim 10^{-11}$M$_{\odot}$/yr (see Methods). During the high mode, and removing the contribution of the low-state flux, we estimate a mass accretion rate of $\dot{M}_{WD} \gtrsim 2 \times 10^{-10}$M$_\odot$/yr. 

The rapid moding timescale and large luminosity drop both indicate that the origin is associated with a sudden reduction in the mass accretion rate onto the white dwarf. Changes from high to low states and vice versa have been observed in the so-called novalikes, including magnetic systems, as well as in VY Scl stars, but occurring on much longer (days to months) timescales where the mass transfer rate from the companion almost, or totally, turns off$^{14,19,20,22}$. The rapid transitions displayed in TW Pic have never been observed in an accreting white dwarf, suggesting a different mechanism regulating the mass accretion rate between the two modes. A plausible scenario may be that, in the high mode, the disk material reaches the white dwarf surface and is accreted. Suddenly and abruptly, the mass transfer rate through the inner-disk regions is substantially reduced during the mode switch. The spinning white dwarf magnetosphere is then able to prevent material from reaching the white dwarf surface, resulting in a substantial luminosity drop. During this mode of low mass transfer rate the spinning magnetosphere is able to regulate the inflow of residual disk material onto the white dwarf surface, which results in the observed magnetically gated bursts. Once the disk mass transfer rate is enhanced again, the system switches to the high luminosity mode. 

The mode switching timescale may be associated with the time matter takes to empty a reservoir region within the magnetospheric radius. In practice this is related to the viscous timescale in the region, $t_{visc} \approx \left( \frac{t_{dyn}}{\alpha (H/R)^2} \right)$, where $t_{dyn}=\left( \frac{4 \pi^2 R^3}{GM} \right)^{1/2}$~is the dynamical (Keplerial) timescale at disc radius $R$, with $M$ being the mass of the accretor and $G$ the Gravitational constant. In the above, $\alpha$ is the dimensionless viscosity parameter and $H$ the disk height of an $\alpha$-disk model$^{23}$. As TW Pic displays magnetically gated bursts in the low mode, we can set the disc magnetospheric truncation radius at co-rotation$^{3,21}$ $R_{mag} \approx R_{co} = \left( \frac{G M_{WD}P_{s}^2}{4 \pi^2} \right)^{1/3}$. To evaluate the viscous timescale in a typical accreting white dwarf, we adopt  a white dwarf mass$^{24}$ of 0.8M$_{\odot}$ with radius$^{25}$ $R_{WD} = 6.89 \times 10^8$ cm. For white dwarfs with spin periods in the range $P_s$= 13s - 1000s, the truncation radius occurs in the range $R_{tr}\approx 1R_{WD}$ - $6R_{WD}$. Using the inferred disk mass transferred rate of $\dot{M} \gtrsim 2 \times 10^{-10}$M$_\odot$/yr and $R_{mag}\approx R_{co}$, the resulting magnetic field strength on the white dwarf surface can be constrained to $B\lesssim 10^6$ Gauss, where the exact value would depend on the white dwarf spin period.  In this case the viscous timescale to drain the disk from the truncation radius lies in the range $t_{visc} \approx$20 min - 1 day. Although the resulting range is somewhat large, it is still consistent with the observed mode switching timescale observed in TW Pic. The viscous timescale is calculated assuming $\alpha (H/R)^2 \approx 0.01$. Theoretical models of accretion disks around white dwarfs predict a lower value for both $H/R$ and $\alpha$ but several observational results$^{26,27}$ suggest a much higher value for both parameters. Thus both $\alpha$ and $(H/R)$ are somewhat arbitrary, as long as they are not greater than 1. 

This simple interpretation may also apply to the few observed transitional millisecond pulsars (tMSPs) which were identified to show mode switching behaviour at X-ray and optical wavelengths, but on a faster, $\approx$tens of seconds, timescales and with much shorter duration of the low modes$^{10,28,29}$. Using standard parameters for these fast spinning neutron stars (M$_{NS}\approx 1.4$M$_{\odot}$, $B_{NS}\approx 10^8$-$10^9$ G, and mass accretion rate of $\dot{M}_{NS}\approx 10^{-12}$M$_\odot$/yr), the magnetospheric radius where the disc is truncated is $R_{tr}\approx$100km-400km. Adopting a similar value of $\alpha (H/R)^2 \approx 0.01$ yields viscous timescales in the range $t_{visc}\approx$2s-12s, also well-matching with what is observed in tMSPs. It is worth noting that the high-to-low mode switching in the most studied tMSP PSRJ1023$+$0038 generally occur faster than the opposite low-to-high mode switching$^{30}$, and that the duration of low modes is variable with time. This also provides a strong analogy with the mode switching seen in TW Pic. Another possibility could be that what precipitates these episodic mode transitions is an abrupt reconfiguration of the white dwarf magnetosphere between discrete metastable states, similarly to what has been suggested$^{30,31}$ for PSRJ1023$+$0038. In either case whether the transitions are induced by mass accretion rate variations or reconfigurations of the white dwarf magnetosphere, TW Pic may be exhibiting transitions from an accretion to a propellor/ejection phase during the moding behaviour similar to what is observed in tMSPs. Nonetheless, the observation of rapid and transitional mode switching observed in TW Pic appears to be analogous to what is observed in the tMSPs only so far. TW Pic is thus far unique in its properties, but further high-cadence and long monitoring observations as those provided by \tess\ may reveal more of these transitional accreting white dwarfs (tAWDs).

\section*{METHODS}

\subsubsection*{Data Sources}
The \tess\ data for TW Pic was obtained from The Barbara A. Mikulski Archive for Space Telescopes (MAST) in reduced and calibrated format$^{32,33}$. The \tess\ telescope/detector combination is sensitive to light across a wide range of wavelengths (600 nm - 1000 nm). Although this wide passband maximizes the signal-to-noise ratio and gives robust relative brightnesses over time for sources, it makes it difficult to calibrate the \tess\ photometry, either in absolute terms or against other observations obtained in standard (narrower) passbands. In the case of TW Pic, we have access to ASAS-SN$^{17,18}$ ground based V and g band photometry. This data set includes g-band observations spanning nearly 3 years, including the entire period over which \tess\ observed the source. We have used this overlap to establish an approximate transformation of the \tess\ count rates for TW Pic into standard g-band magnitudes (see Extended Data Fig.~3). To achieve this, we converted the ASAS-SN g-band magnitudes into instrumental fluxes, then selected the data points between \tess\ and ASAS-SN that were taken within 0.002 days (173s) of each other. These selected data points were correlated and linearly fitted. The resulting correlation allows the \tess\ count rates to be converted into equivalent ASAS-SN g-band magnitude.

\subsubsection*{Accretion Rate}
We can take the luminosity change of $\approx 3.5$ during the mode switching to represent a change in mass accretion rate onto the white dwarf by at least the same amount. In practice this assumes that the inner disk radius remains unchanged during the mode transitions, and that there is no bolometric correction to be applied to the luminosity variations. Thus the inferred change in mass accretion rate of $\approx 3.5$ sets a hard lower limit to the true mass accretion rate change. We can nonetheless infer absolute values for the mass accretion rates during both modes, with the caveat that these must also be considered lower limits on the true values. During the low modes TW Pic displays quasi-periodic bursts, which we interpret as a signature of magnetically gated accretion. The minimum count rate between the bursts is approximately constant, at a level corresponding to $m_g \approx 16.9$. This stable count rate is therefore most likely due to a combination of light from the white dwarf, the donor star and the accretion flow at $R>R_{co}$. In order to estimate the time-averaged count rate that is released by the bursts themselves, we calculate the average count rate across the entire low mode and then subtract the stable minimum count rate. We then use our ASAS-SN-based calibration to convert the time-averaged burst count rate into an equivalent g-band flux density($f_g \approx 3 \times 10^{-15}$erg s$^{-1}$cm$^{-2}$\AA$^{-1}$). This, in turn, allows us to estimate the corresponding g-band luminosity via $L_g\approx 4 \pi d^2 f_g \lambda_{eff,g} \approx 3 \times 10^{32}$erg s$^{-1}$, where $d=428$ pc is the distance towards TW Pic based on the Gaia eDR3 parallax measurement$^{34}$, and $\lambda_{eff,g}=4754$\AA\ is the effective wavelength of the g-band filter. In the absence of information about the spectral shape of the radiation produced by the bursts, we assume that most of this radiation emerges in the optical region and estimate the burst-related accretion luminosity as $L_{acc}\approx L_g$, i.e. without applying a bolometric correction. We then convert this luminosity into an estimate of the accretion rate onto the white dwarf via $L_{acc}\approx G \dot{M} M_{WD} (1/R_{WD} - 1/R_{in}$) under the assumption that $R_{in}>>R_{WD}$. This estimate assumes that $L_{acc}$ represents the gravitational potential energy release associated with material falling from $R_{in}$ (the inner edge of the truncated accretion disk) to $R_{WD}$ (the surface of the WD). In the limit $R_{in}>>R_{WD}$, this yields $\dot{M} \gtrsim 10^{-11}$M$_{\odot}$yr$^{-1}$. Here, we have assumed$^{24,25}$ $M_{WD}=0.8$M$_{\odot}$ and $R_{WD}=6.89\times 10^8$cm. Our estimates of both $L_{acc}$ and $M$ are approximate lower limits, since we have not made any bolometric correction. If the radiating region is hot, for example, most of the magnetically gated burst energy may be released in the far-ultraviolet or X-ray band, rather than in the optical$^{35-38}$. To estimate the mass accretion rate in the high mode we use the high mode g-band magnitude of $m_g\approx 14.2$ and convert this to $L_g\approx 9 \times 10^{32}$erg s$^{-1}$. Assuming the disk now extends to the white dwarf surface and again that $L_{acc} \approx L_g$, we can estimate the mass accretion rate using $L_{acc} \approx G \dot{M} M_{WD} (1/R_{WD})$, yielding  $\dot{M} \gtrsim 2 \times 10^{-10}$M$_{\odot}$yr$^{-1}$. If we instead assume a standard $\alpha$-disk model$^{23}$, we are able to reproduce the $m_g\approx 14.2$ observation in the high mode adopting a disk mass transfer rate of $\dot{M} \approx 2 \times 10^{-9}$M$_{\odot}$yr$^{-1}$ and assuming the disk extends from a circularisation radius (calculated assuming a system orbital period of 6.5 h and mass ratio $q$=0.45) up to the white dwarf surface.

\subsubsection*{Power Spectra}
Lomb-Scargle periodograms$^{39,40}$ of TW Pic have been computed using \tess\ Cycle 1 data and both the low and high modes observed during Cycle 3. The periodogram for Cycle 1 data has been computed using the available 120-s cadence data, while both the low and high modes observed in Cycle 3 use 20-s cadence data. Overall, the periodograms do not reveal any strong coherent signal, with the possible exception of a $P\approx 6.956$ hour signal detected in Cycle 1 data when TW Pic was in a high accretion state. We determine the significance level of this signal by computing the full periodogram across the entire Cycle 1 data. Ideally we require knowledge of the underlying broad-band noise variability to compute significance levels. As this is not precisely known (and may be time variable) we instead smooth the computed periodogram using a running mean (1000 data points) in an attempt to model the broad-band variability. This smoothed periodogram is then used to simulate 100,000 lightcurves$^{41}$. The simulated light curves are then interpolated onto the \tess\ timestamps. For each we again compute the periodogram and record the highest power attained in any frequency interval. Extended Data Fig.~1 shows how the $P\approx 6.956$h appears to be significantly detected above the 1 in 100,000 level. We have performed the same analysis using 20-s cadence data from both the low and high mode separately identified in Figure~1. It is interesting to note that a similar signal appears to be relatively strong at $P\approx 6.964$h. We note that the consequence of our methodology to infer the underlying broad-band noise component may lead to an overestimate of the variability power at a given frequency, making the significance levels less conservative than they may appear. It is worth noting that signals of  $\approx 6.06$ h and $\approx 2.1$ h have been previously reported for TW Pic in optical spectroscopy and photometry$^{15,16}$, although not further confirmed$^{12}$. Our interpretation is that the former may be related to the $P \approx 6.964$ h detected in the \tess\ data. This may be a beat (also referred to as a superhump) between a super orbital signal (either from the precession of an elliptical or a tilted disk) and the system orbital period. On the other hand the $\approx 2.1$ h  reported signal in the literature may be related to the magnetic gating timescale observed in TW Pic in the low mode, although we are unaware about the optical level of the system when it was detected$^{15,16}$.

\subsubsection*{Archival X-ray and optical observations of TW Pic}
During the low modes observed by \tess, TW Pic was at an optical level of $m_g \approx 16.5$. The faintest reported optical magnitude of TW Pic was $m_B \approx 17.9$, $m_V \approx 18.1$, $m_R \approx 17.9$ and $m_I \approx 17.0$ on August 12-13 1990$^{12}$ . The observed colours and derived absolute magnitudes using the Gaia eDR3 distance of 428pc$^{33}$ would indicate the contributions of a hot component, compatible with a $\approx$20,000K white dwarf, and of a cooler component, compatible with a K3-K7V star. This suggests that the optical emission of TW Pic, at that epoch, was due to the bare white dwarf and secondary star photospheres and hence that accretion was turned off. Furthermore, the \textit{ROSAT} All Sky survey observed TW Pic between Aug. 12-22, 1990 and between Feb. 13-15, 1991. From inspection of the X-ray light curve of TW Pic, provided in the Second \textit{ROSAT} X-ray Source (2RXS) catalogue products$^{42}$, the source is not significantly detected in August, while it is detected at the $4\sigma$ level in February, further supporting that accretion switched off in August 1990. Therefore, the difference of $\approx$1.5 magnitudes between the deep low mode observed by \tess\ in Cycle 3 and the faintest level attained in August 1990 implies that during the deep low modes the disk has not vanished entirely despite a substantial decrease in the mass accretion rate. This is also supported by the presence of magnetically gated bursts observed in the low modes of TW Pic. No X-ray data are available during the Cycle 3 low modes for further support.

\bibliography{scibib}

\bibliographystyle{Science}

\paragraph*{Acknowledgments:}
This paper includes data collected by the \tess\ mission. Funding for the \tess\ mission is provided by the NASA's Science Mission Directorate. Some of the data presented in this paper were obtained from the Mikulski Archive for Space Telescopes (MAST). STScI is operated by the Association of Universities for Research in Astronomy, Inc., under NASA contract NAS5-26555. Support for MAST for non-HST data is provided by the NASA Office of Space Science via grant NNX09AF08G and by other grants and contracts. This paper uses data from the ASAS-SN project run by the Ohio State University. D.dM. and A.P.  acknowledge financial support from the Italian Space Agency (ASI) and National Institute for Astrophysics (INAF) under agreements ASI-INAF I/037/12/0 and ASI-INAF n.2017- 14-H.0 and from INAF ‘Sostegno alla ricerca scientifica main streams dell’INAF’, Presidential Decree 43/2018 and from INAF ‘SKA/CTA projects’, Presidential Decree 70/2016 and from PHAROS COST Action N. 16214. D.A.H.B acknowledges support from the National Research Foundation. P.J.G. is supported by NRF SARChI grant 111692. We thank the ASAS-SN team for making their data publicly available. We acknowledge T. Boller and F. Haberl for providing the \textit{ROSAT} lightcurve of TW Pic. This work has also made use of data from the European Space Agency (ESA) mission \textit{Gaia} (https://www.cosmos.esa.int/gaia), processed by the Gaia Data Processing and Analysis Consortium (DPAC, https://www.cosmos.esa.int/web/gaia/dpac/consortium). Funding for the DPAC has been provided by national institutions, in particular the institutions participating in the \textit{Gaia} Multilateral Agreement.

\paragraph*{Author Contributions:}
S.S. analysed the \tess\ data, identified the phenomenon, suggested the analogy to tMSPs, and was the primary author. D.dM. carried out the archival ancillary analysis. P.J.G. carried out the ASAS-SN based calibration of the \tess\ data. D.A.H.B. reviewed previous observations of TW Pic. All authors shared ideas, interpreted the results, commented, and edited the manuscript.

\begin{extFigure*}[h]
\begin{center}
\includegraphics[width = \textwidth]{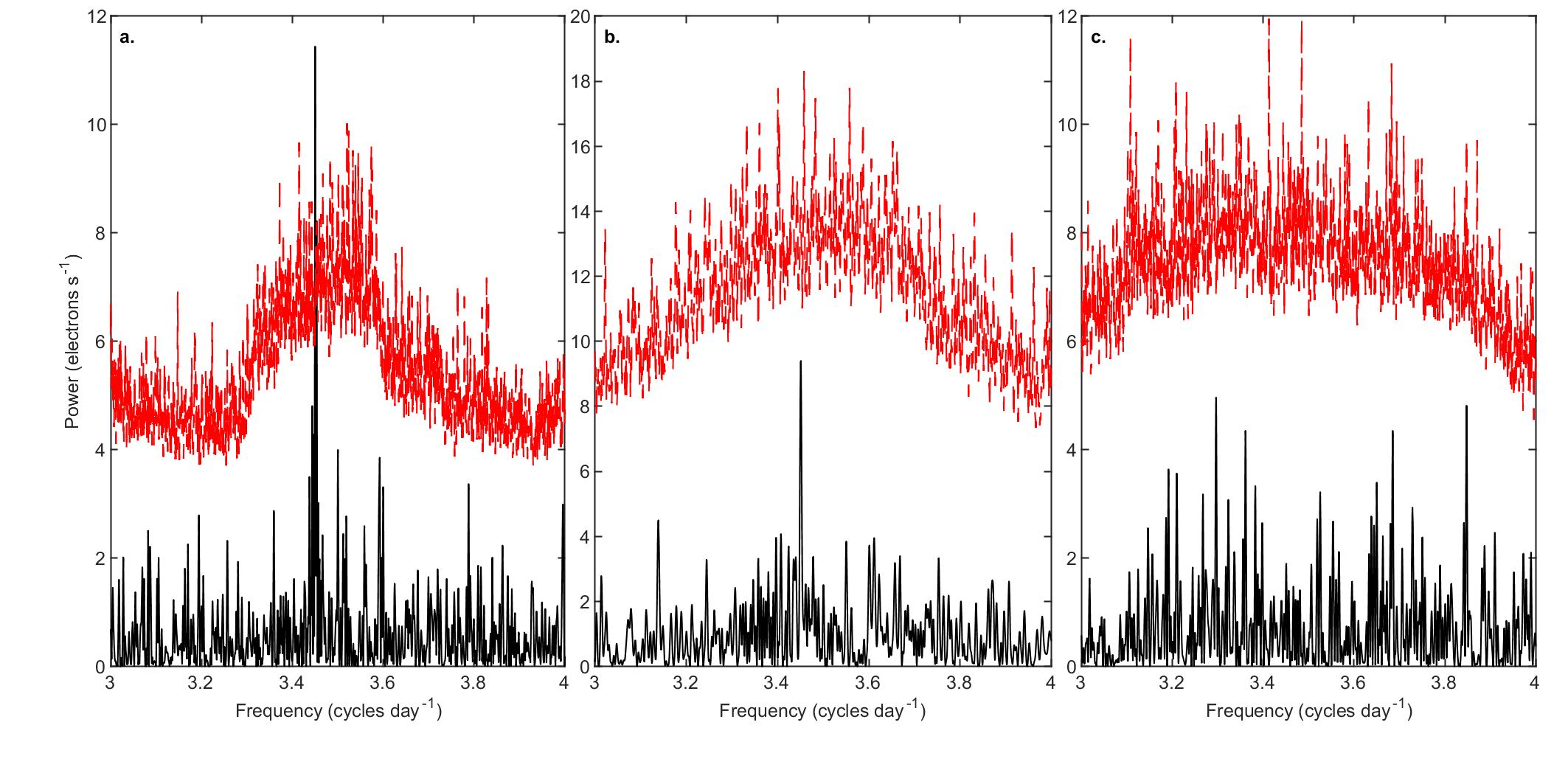}
\caption{
\noindent \textbf{TW Pic power spectra in the 3-4 cycle/day range.}
Lomb-Scargle periodogram using \tess\ data (solid black line) within the 3-4 cycle/day region. \textbf{a.} All Cycle 1 observed in 2 min cadence mode. \textbf{b.} Low mode from Cycle 3 observed at 20 sec cadence. \textbf{c.} High mode from Cycle 3 observed at 20 sec cadence. In all panels the dashed red line represented the 1 in 100,000 significance confidence level.  
}
\end{center}
\end{extFigure*}

\begin{extFigure*}[h]
\begin{center}
\includegraphics[width = \textwidth]{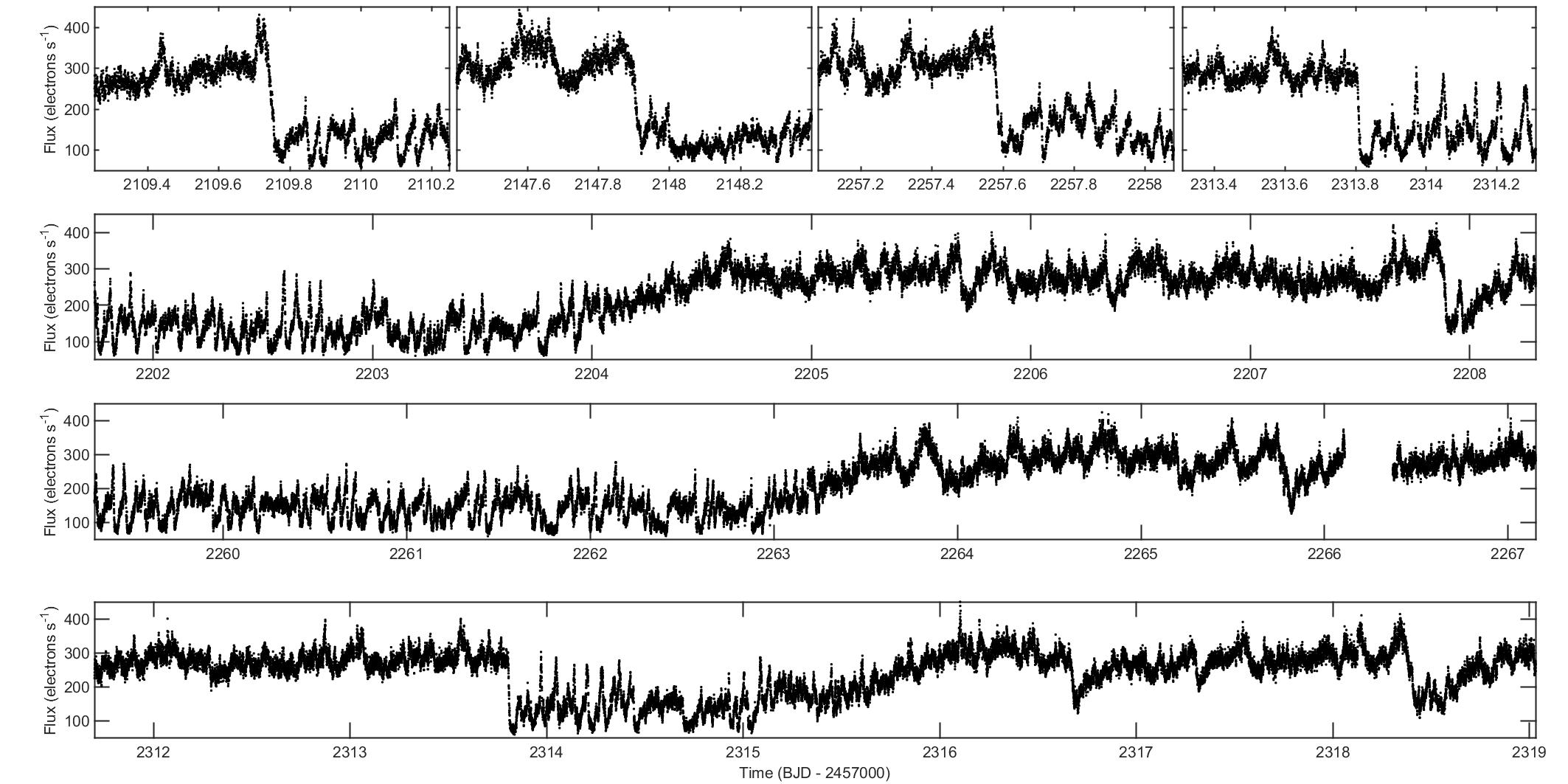}
\caption{
\noindent \textbf{TW Pic mode transition.}
Cycle 3 \tess\ 20 sec cadence lightcurves displaying the four recorded high-to-low mode transitions (top row) and the three recorded low-to-high mode transitions (bottom three rows).  
}
\end{center}
\end{extFigure*}

\begin{extFigure*}[h]
\begin{center}
\includegraphics[width = \textwidth]{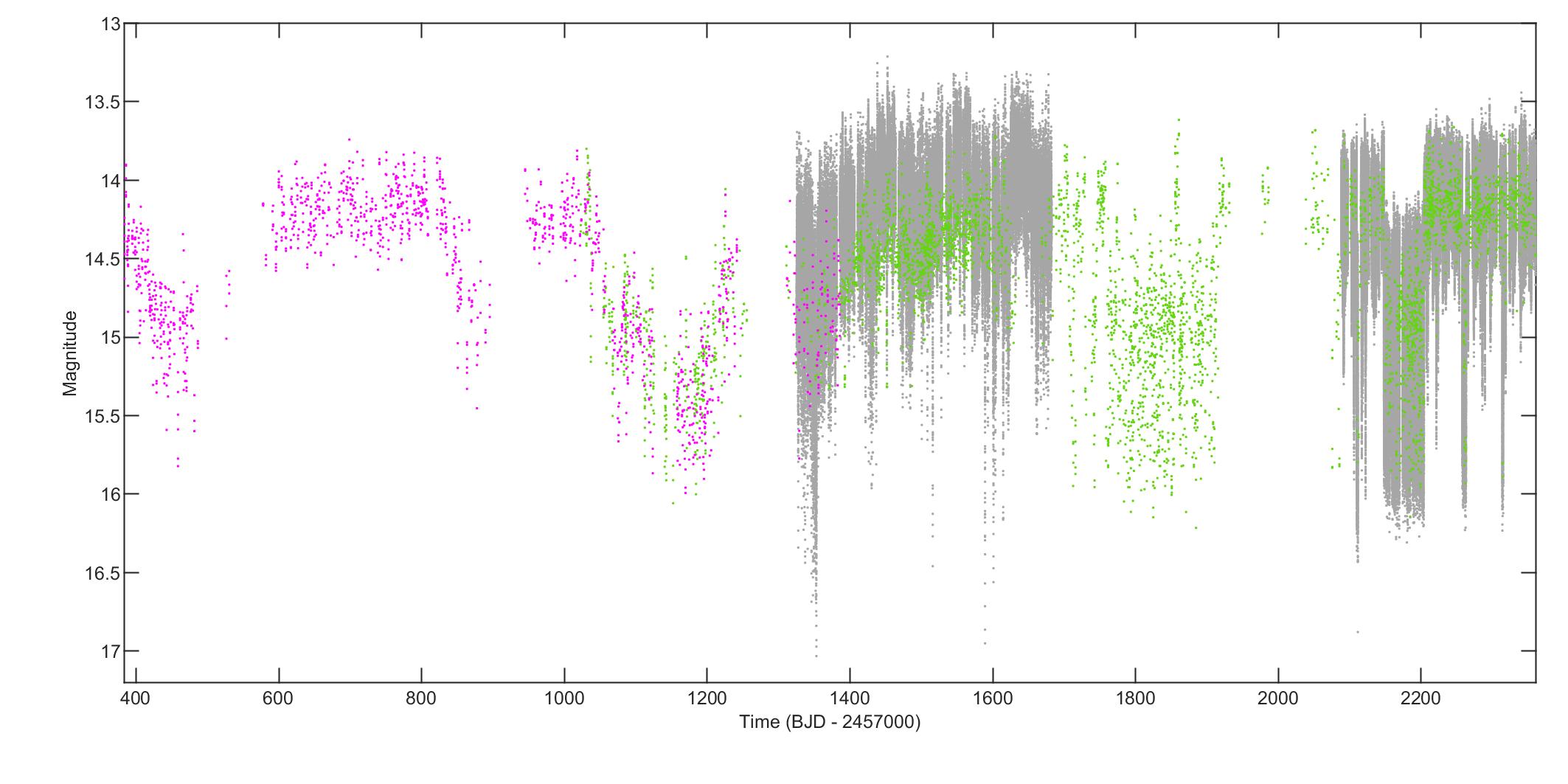}
\caption{
\noindent \textbf{Long term optical variations in TW Pic.}
Cycle 1 and Cycle 3 \tess\ data (grey points) overlaid onto the long term ASAS-SN V-band (magenta points) and g-band (green points) photometry. TW Pic shows evidence for slow ($\approx$100 days) changes in its mass transfer rate from V-band and g-band ASAS-SN observations. The rapid mode transitions are only observed in g-band ASAS-SN data (around day 1800) and during Cycle 3 \tess\ observations.  
}
\end{center}
\end{extFigure*}

\begin{extFigure*}[h]
\begin{center}
\includegraphics[width = \textwidth]{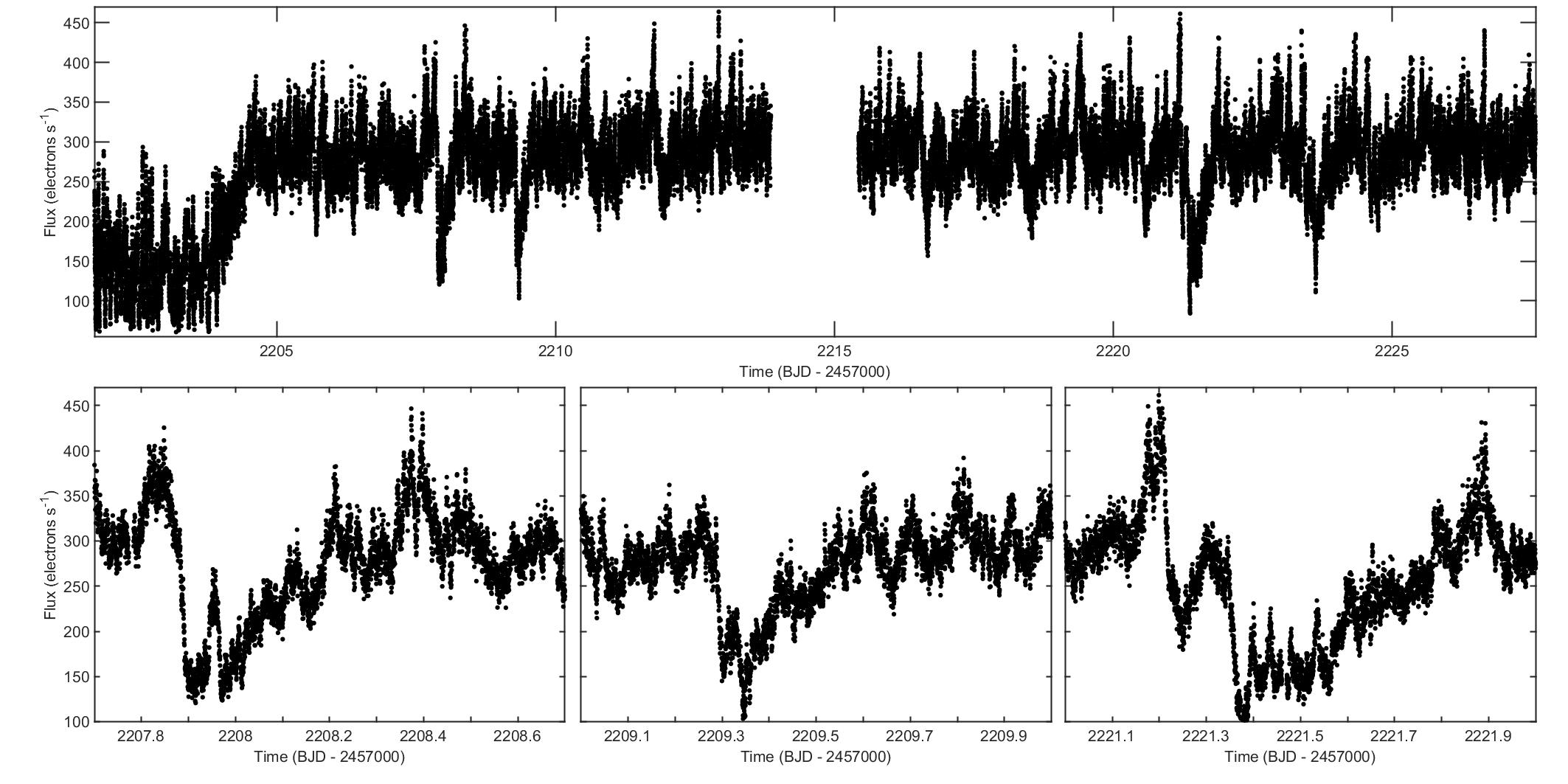}
\caption{
\noindent \textbf{Failed mode transitions in TW Pic.}
Cycle 3 \tess\ 20 sec cadence light curves displaying the low-to-high transition (top panel) as well as enlarged panels (bottom row) showing representative failed mode transitions. 
}
\end{center}
\end{extFigure*}

\end{document}